\documentclass[%
 aip,
 jmp,%
 amsmath,amssymb,
 reprint,%
]{revtex4-2}
\usepackage{graphicx}
\usepackage{epstopdf, epsfig}
\usepackage{dcolumn}
\usepackage{bm}
\usepackage{gensymb}
\usepackage{enumitem}
\usepackage{mathrsfs}
\usepackage[sort&compress]{natbib}
\usepackage{color}
\usepackage{hyperref}

\begin{document}


\title{Exact mathematical formulas for wall-heat flux in compressible turbulent channel flows}
\author{Peng Zhang}
\affiliation{Department of Engineering Mechanics, Zhejiang University, Hangzhou 310027, China}
\author{Yubin Song}
\affiliation{Department of Engineering Mechanics, Zhejiang University, Hangzhou 310027, China}
\author{Zhenhua Xia}
\email{xiazh@zju.edu.cn}
\affiliation{Department of Engineering Mechanics, Zhejiang University, Hangzhou 310027, China}


\date{\today}
\begin{abstract}
In this paper, several exact expressions for the mean heat flux at the wall ($q_w$) for the compressible turbulent channel flows are derived by using the internal energy equation or the total energy equation. Two different routes, including the FIK method and the RD method, can be applied. The direct numerical simulations data of compressible channel flows at different Reynolds and Mach numbers verify the correctness of the derived formulas. Discussions related to the different energy equations, and different routes are carried out, and we may arrive at the conclusion that most of the formulas derived in the present work are just mathematically ones which generally are lack of clear physical interpretation. They can be used to estimate $q_w$, but might not be suitable for explore the underlying physics.
\end{abstract}


\maketitle

\section{Introduction}

Compressible wall-bounded turbulence is an important phenomenon in many industrial and engineering applications~\cite{Liu2010,chen2013}. For example, in high-speed aerospace vehicles, the flows around their surfaces are generally compressible and turbulent, and the drag forces as well as the wall heating rates are two crucial topics \cite{leyva2017relentless,urzay2018supersonic,candler2019rate}. On the one hand, fuel consumption will increase with the drag forces. On the other hand, the large
wall heating rates caused by air friction and shock waves continuously endanger the maneuverability and structural integrity of the vehicles. Therefore, it is very critical to clarify the sources of wall friction and heat flux, and then to explore the possible ways to control them, in addition to the investigations of compressibility effect and the corresponding mechanism on compressible wall-bounded turbulence~\cite{bradshaw1977, Lele1994, Friedrich2007, gatski2013,Pei2013}.

In the past decades, the wall friction was investigated from different aspects \cite{FIK2002,gomez2009,DECK2016,Yoon2016}. Fukagata et al. \cite{FIK2002} firstly derived a simple expression for the skin friction resistance, known as the FIK decompostion, based on the normal integral of momentum equation in incompressible channel, pipe and plane boundary layer flows. For incompressible turbulent channel flows without sidewall blowing or suction, the FIK decomposition can be expressed as:
\begin{equation}
 C_{f,FIK} = \underbrace{ \frac {6} {Re} }_{C_{f1,FIK}} + \underbrace{6 \int_{0}^{1}(1-y)(-\overline {u^{\prime}v^{\prime}})dy}_{C_{f2,FIK}},
 \label{eq:cf1}
\end{equation}
where $C_f=\nu(\partial \langle u \rangle/\partial y)_w/(u_b^2/2)$ is the skin friction coefficient and $Re=u_bh/\nu$ is the Reynolds number with $\nu$ being the kinematic viscosity, $h$ being the channel half height and $u_b$ being the mean bulk velocity. In the above formula, $y$ and $\overline {u'v'}$ are normalized by $h$ and $u_b^2$. It is easy to see that the skin friction coefficient $C_f$ can be decomposed into two parts, including a laminar contribution term $C_{f1,FIK}$, of which the value exactly equals to the laminar solution, and a turbulent contribution term $C_{f2,FIK}$, whose value is proportional to the linearly weighted average of the Reynolds shear stress. The authors believed that the formula enables quantitative discussions on the drag reduction or enhancement effects brought about the modification of the flow structures, and that it might serve as a clue for the development and evaluation of new control strategies. Gomez et al.~\cite{gomez2009} extended the incompressible FIK decomposition to compressible wall-bounded flows, and it was reported that two additional compressibility-induced terms, namely the compressible contribution term and the compressible-turbulent interaction term, will come out besides the laminar and turbulent terms in incompressible channel flows. Their numerical results also showed that the main contribution to the skin friction in compressible channel flows comes from the turbulent contribution term even at Mach number $Ma=2$. It should be noted that the FIK decomposition is an exact mathematical formula, and thus it can be used to estimate the skin friction coefficients. Following the idea of the FIK decomposition, Mehdi and his collaborators~\cite{Mehdi2011,Mehdi2014} proposed an integral formula to estimate the skin friction coefficient in wall-bounded turbulence by using the experimental data, and Xia et al.~\cite{xia2020} further derived the skin friction coefficient formulas for incompressible turbulent channel flows and turbulent boundary layer flows with zero pressure gradient in terms of the mean velocity and the Reynolds shear stress in an arbitrary wall-normal region.

Renard and Deck~\cite{DECK2016} derived another theoretical decomposition of the mean skin friction for incompressible wall-bounded turbulence, known as the RD decomposition, from a mean streamwise kinetic-energy budget in an absolute reference frame, in which the undisturbed fluid is not moving. For incompressible turbulent channel flows, the skin friction coefficient can be decomposed into a diffusion term and a dissipation term as
\begin{equation}
C_{{f}, {RD}}=\underbrace{\int_{0}^{1} \frac{2}{R e}\left(\frac{\partial\langle u\rangle}{\partial y}\right)^{2} {~d} y}_{C_{{f} 1, {RD}}}+\underbrace{\int_{0}^{1} 2\left(-\left\langle u^{\prime} v^{\prime}\right\rangle\right) \frac{\partial\langle u\rangle}{\partial y} {~d} y}_{C_{{f} 2, {RD}}}.
 \label{eq:cf2}
\end{equation}
Here, $y$, $\langle u\rangle$ and $\overline {u'v'}$ are normalized by $h$, $u_b$ and $u_b^2$ respectively, and $C_f$ can be viewed as the mean power supplied in the $u_b$-moving frame by the wall to the fluid through skin friction, and it is transformed into heat by direct dissipation $C_{{f} 1, {RD}}$ and into the turbulent kinetic energy by the production $C_{{f2}, {RD}}$. It is argued by the authors that the RD decomposition, relying on the energy budget, had a straightforward physical interpretation and can be useful to design new drag reduction strategies~\cite{DECK2016}. The RD decomposition was also extended to compressible wall-bounded turbulent flows~\cite{weipeng2019,Fan2019} to investigate the effect of Reynolds number and scalings of each term.

For heat flux at the wall, Fukagata et al.~\cite{FIK-Heat2005} used to derive a mathematical relation of the contribution of turbulent heat flux to the Nusselt number for incompressible channel with isothermal conditions. 
Similar formula was derived for pipe flows at low Mach numbers by Liu et al.~\cite{liu2020}. However, in these works the temperature was treated as a passive scalar, which is not the case for fully compressible wall-bounded flows. Although there are many works on the wall-heat flux in compressible wall-bounded turbulent flows, the heat transfer coefficients were generally estimated directly through the wall-normal gradient of the mean temperature at the wall~\cite{Morinishi2004,Duan2010,zhang2012}. There were also some trials to relate the wall heat transfer to the skin-friction. Hopkins and Inouye~\cite{HOPKINS1971} used to introduce an empirical Reynolds analogy factor to relate the heat transfer at the wall to the skin-friction on flat plates, and reported that the factor of 1.0 should be used for hypersonic flows while a value of 1.2 was recommended near adiabatic surfaces exposed to supersonic and low Mach numbers based on the experimental data. For fully developed compressible turbulent channel flows with symmetric isothermal boundary conditions, Huang et al.~\cite{huang1995} reported
\begin{equation}
q_w\equiv-\lambda \frac{d\langle T\rangle}{dy}|_w=-u_b\tau_w,\label{eq:Huang-b}
\end{equation}
which is based on the overall energy balance argument that the total pressure
work done, i.e. the total heat generation, across the channel and the heat transfer into the walls are the same. Here, $q_w$ is the heat flux at the wall, $\lambda$ is the thermal conductivity, $\tau_w$ is the wall shear stress, and $u_b=\int_0^h\langle \rho u\rangle dy/\int_0^h\langle \rho \rangle dy$ is the mean bulk velocity in a compressible symmetric channel. This formula was also discovered by Ghosh et al.~\cite{Ghosh2010} and Li et al.~\cite{weipeng2019}. It is easy to see that all the above works focused on the estimation of the heat transfer at the wall.

Recently, Zhang and Xia \cite{zhangpeng2020} derived a mathematical expression to relate the wall heat flux to the integral of the turbulent statistics in the whole channel following a similar procedure as the FIK decomposition for the skin friction, aiming to explore the source of wall heat flux in compressible channel flows. Their formula was based on the internal energy equation and the numerical result showed that the contribution from the viscous stress work dominates the main contributions (around 90\%) to wall heat flux. More recently, another formula on the wall heat flux was derived by Sun et al.~\cite{Sun2021} for the compressible boundary layer following the idea of the RD decomposition. The formula was started with the total energy equation, and their numerical results showed that Reynolds stress transport term contributes a lot to $q_w$ which is in sharp contrast to the results obtained in Zhang and Xia \cite{zhangpeng2020}.

From the above discussions, it is easy to see that different formulas for the wall heat flux can be derived based on different procedures and different equations. However, the intrinsic connections among different formulas have not been discussed yet. In this paper, we will derive four different formulas for the wall heat flux from the perspectives of total energy equation and internal energy equation by following the FIK and RD methods, and the formulas will be compared with each other to explore the underlying connections.

\section{Exact mathematical formulas for the wall heat flux $q_w$}\label{sec:Derivation}

In this section, we are going to derive the exact mathematical formulas for the wall heat flux $q_w$ in a fully developed compressible turbulent channel flow, where the fluid between two parallel planes is driven by a time-varying uniform external force $f$ along the streamwise direction. In the following, $u_1, u_2, u_3$ denote the streamwise ($x_1$), wall-normal ($x_2$) and spanwise ($x_3$) velocity components, respectively; $u, v, w$ and $x, y, z$ are used interchangeably with $u_1, u_2, u_3$ and $x_1, x_2, x_3$, respectively. For this problem, the equation for internal energy ($e=\rho C_vT$) ($C_v$ is the specific heat at constant volume) reads as
\begin{equation}
 \frac{\partial e}{\partial t}+\frac{\partial\left(e u_{k}\right)}{\partial x_{k}}+\frac{\partial q_{k}}{\partial x_{k}}=-p \frac{\partial u_{k}}{\partial x_{k}}+\tau_{ik} \frac{\partial u_{i}}{\partial x_{k}},
 \label{eq:e}
\end{equation}
and the equation for total energy ($E=e+\rho u_iu_i/2$) is
\begin{equation}
 \frac{\partial E}{\partial t}+\frac{\partial\left(E u_{k}\right)}{\partial x_{k}}+\frac{\partial q_{k}}{\partial x_{k}}=- \frac{\partial p u_{k}}{\partial x_{k}}+ \frac{\partial \tau_{ik} u_{i}}{\partial x_{k}}+f u_1,
 \label{eq:E}
\end{equation}
where
\begin{equation*}
\tau_{ij}=\mu\left(\frac{\partial u_i}{\partial x_j} + \frac{\partial u_j}{\partial x_i} -\frac{2}{3} \frac{\partial u_k}{\partial x_k}\delta_{ij}\right),\qquad q_k=-\lambda \frac{\partial T}{\partial x_k},
\end{equation*}
are the viscous stress tensor and the conductive heat flux, respectively. $T$ is the temperature and $p$ is the pressure. It is easy to know that the pressure dilatation term $P_d=-p {\partial u_{k}}/{\partial x_{k}}$ and the viscous action term $V_a=\tau_{ik}{\partial u_{i}}/{\partial x_{k}}$ are the interaction terms between the kinetic and internal energy~\cite{Mittal2019}.

Both the Reynolds averaging and the Favre averaging will be used in the following, where $\bar{\phi}=\langle\phi\rangle$ denotes the Reynolds averaging, while $\widetilde{\phi}=\{\phi\}= \overline{\rho\phi}/\overline{\rho}$ denotes the density-weighted (Favre) averaging operator. The corresponding fluctuations are denoted as $\phi'=\phi-\overline{\phi}$ and $\phi''=\phi-\widetilde{\phi}$ respectively. Equations~\eqref{eq:e} and ~\eqref{eq:E} can be averaged using the Reynolds averaging operator to obtain the Reynolds-averaged equations for the internal energy and total energy as
\begin{equation}
 \frac{\partial \bar{e}}{\partial t}+\frac{\partial\left(\overline{e u_{k}}\right)}{\partial x_{k}}+\frac{\partial \overline{q_{k}}}{\partial x_{k}}=\overline{-p \frac{\partial u_{k}}{\partial x_{k}}}+\overline{\tau_{ik} \frac{\partial u_{i}}{\partial x_{k}}},
 \label{eq:e-mean}
\end{equation}
and
\begin{equation}
 \frac{\partial \bar{E}}{\partial t}+\frac{\partial\left(\overline{E u_{k}}\right)}{\partial x_{k}}+\frac{\partial \overline{q_{k}}}{\partial x_{k}}=- \frac{\partial \overline{p u_{k}}}{\partial x_{k}}+\frac{\partial \overline{\tau_{ik} u_{i}}}{\partial x_{k}}+\overline{fu_1}.
 \label{eq:E-mean}
\end{equation}

For a fully developed turbulent channel flow, the Reynolds averaging operator can be substituted by the averaging in $x$, $z$ and $t$, and thus
\begin{equation*}
\frac{\partial \bar{\phi}}{\partial x}=\frac{\partial \bar{\phi}}{\partial z}=\frac{\partial \bar{\phi}}{\partial t}=0
\end{equation*}
for any variable $\phi$. The above Reynolds averaged equations~\eqref{eq:e-mean} and~\eqref{eq:E-mean} can be further simplified as
\begin{equation}
 \frac{\partial\left(\overline{e v}\right)}{\partial y}+\frac{\partial \overline{q_{y}}}{\partial y}=\overline{P_d}+\overline{V_a},
 \label{eq:e-mean-s}
\end{equation}
and
\begin{equation}
\frac{\partial\left(\overline{E v}\right)}{\partial y}+\frac{\partial \overline{q_y}}{\partial y}=- \frac{\partial \overline{p v}}{\partial y}+\frac{\partial \overline{\tau_{i2} u_{i}}}{\partial y}+\overline{fu}.
 \label{eq:E-mean-s}
\end{equation}

\subsection{Exact mathematical formulas for $q_w$ by the FIK method}\label{sec:qw_FIK}

The basic idea of the FIK method is the normal integration of the balance equation, and it can be applied to the energy equation. By integrating equations~\eqref{eq:e-mean-s} and~\eqref{eq:E-mean-s} from 0 to $y$, we may obtain the balance equations for $q_w$ as

\begin{equation}
 q_w = \overline{q_{y}}(y) + \overline{e v}(y) - \int_0^y\overline{P_d}(y_1)dy_1 - \int_0^y\overline{V_a}(y_1)dy_1 - \overline{e v}|_w,
 \label{eq:qw-em1}
\end{equation}
and
\begin{eqnarray}
q_w &=& \overline{q_{y}}(y) + \overline{E v}(y) + \overline{p v}(y) - \overline{\tau_{i2} u_{i}}(y) - \int_0^y\overline{fu}(y_1)dy_1\qquad\nonumber \\
& &\qquad - \overline{E v}|_w - \overline{p v}|_w + \overline{\tau_{i2} u_{i}}|_w.
 \label{eq:qw-Em1}
\end{eqnarray}

Equation~\eqref{eq:qw-em1} was previously reported by Ghosh et al.~\cite{Ghosh2010} and Zhang and Xia~\cite{zhangpeng2020}. Theoretically speaking, equations~\eqref{eq:qw-em1} and~\eqref{eq:qw-Em1} can be used to estimate $q_w$. However, there might be considerable errors coming from the imperfect balance of the equations due to the insufficient sample sizes (see similar results for the skin-friction by Xia et al.~\cite{xia2020}). 

One way to overcome the drawback of the imperfect balance of equations~\eqref{eq:qw-em1} and~\eqref{eq:qw-Em1} is to integrate them one more time from 0 to a certain height $\delta$ in the wall-normal direction. Applying the equality
\begin{equation}
\int_{0}^{\delta} \int _{0}^ {y}\beta(y_1) dy_1 dy=\int_{0}^{\delta} (\delta-y)\beta(y) dy
\label{eq:jifen}
\end{equation}
for an arbitrary physical quantity $\beta$, we may arrive at the following formulas for $q_w$:
\begin{eqnarray}
q_w&=&\underbrace{\frac{1}{\delta}\int_0^\delta\overline{q_y}dy}_{MH}+\underbrace{\frac{1}{\delta}\int_0^\delta C_v\overline{\rho T''v''}dy}_{TH}\nonumber \\
&+&\underbrace{\frac{1}{\delta}\int_0^{\delta}(y-\delta)\overline{P_d} dy}_{PW} +\underbrace{\frac{1}{\delta}\int_0^{\delta} (y-\delta)\overline{V_a}dy}_{VW} ,
 \label{eq:qw-e-mean}
\end{eqnarray}
and
\begin{eqnarray}
q_w&=&\underbrace{\frac{1}{\delta}\int_0^\delta\overline{q_y}dy}_{MH}+\underbrace{\frac{1}{\delta}\int_0^\delta C_v\overline{\rho T''v''}dy}_{TH}\nonumber \\
&+&\underbrace{\frac{1}{\delta}\int_0^{\delta}\overline{pv} dy}_{PWT}+\underbrace{\frac{1}{\delta}\int_0^{\delta} -\overline{u_i \tau_{i2} }dy}_{VWT}+
\underbrace{\frac{1}{\delta}\int_0^{\delta} \overline{\rho v'' u_{k}''u_{k}''/2}dy}_{TTH} \nonumber\\
&+&\underbrace{\frac{1}{\delta}\int_0^{\delta} \widetilde{{u}_{k}} \overline{{\rho} {v}'' {u}_{{k}}''} dy}_{RSH}+
\underbrace{\frac{1}{\delta}\int_0^{\delta} (\delta-y) (-\overline{fu})dy}_{EW}.
 \label{eq:qw-E-mean}
\end{eqnarray}
Here, the no-slip boundary conditions are assumed for all three velocity components at the wall, and thus the terms at the wall at the right-hand side of equations~\eqref{eq:qw-em1} and~\eqref{eq:qw-Em1} are zero. If we would like to include the blowing or suction at the wall, another term related to the blowing or suction will exist straightforwardly~\cite{zhangpeng2020}.

Equation \eqref{eq:qw-e-mean} shows that the heat flux at the wall can be decomposed into four terms: the contribution from the turbulent heat transfer (TH), the contribution from the molecular heat transfer (MH), the contribution from the pressure work (PW), the contribution from the work of viscous stress (VW). The MH and TH terms are the volume-averaged of the molecular heat transfer and the turbulent heat transfer in the wall normal direction respectively, while the PW and VW terms are the weighted average of the pressure dilatation and the viscous action terms respectively, where the weight is linearly decreasing with the distance from the wall~\cite{zhangpeng2020}. According to equation~\eqref{eq:qw-E-mean}, $q_w$ can also be decomposed into seven terms if the total energy equation is used. Besides the TH and MH terms, the two terms PWT and VWT denote the pressure and viscous stress related work respectively, and three terms, including the contribution from the turbulent transport (TTH), the contribution from the Reynolds stress transport (RSH), the contribution from the  external-force work (EW), are new ones.

From a mathematical point, the height $\delta$ in equations~\eqref{eq:qw-e-mean} and \eqref{eq:qw-E-mean} are rather arbitrary, and a different value will deliver a different message. Nevertheless, for compressible turbulent channel flows with symmetric boundary conditions, the half-channel integral is unique, i.e. $\delta=h$, where the information across the whole channel was used. In the following discussions, $\delta=h$ is assumed unless otherwise stated.

\subsection{Exact mathematical formulas for $q_w$ by the RD method}\label{sec:qw_RD}

Following the short route of the RD decomposition for the mean skin friction, exact mathematical formulas for $q_w$ can also be obtained. By multiplying both sides of equations~\eqref{eq:e-mean-s} and \eqref{eq:E-mean-s} by $(\widetilde{u}-u_b)$, and then integrating them over the half channel once, we may arrive at

 \begin{eqnarray}
q_w&=&\underbrace{\frac{1}{u_b}\int _{0}^{h} \overline{q_y}\frac{\partial \widetilde{u}}{\partial y}dy}_{MH_{RD}}+\underbrace{\frac{1}{u_b}\int _{0}^{h}C_v\overline{\rho T^{ \prime \prime} v^{ \prime \prime}}\frac{\partial \widetilde{u}}{\partial y}dy}_{TH_{RD}}\nonumber \\
&+&\underbrace{\frac{1}{u_b}\int _{0}^{h}(\widetilde{u}-u_b)\overline{P_d} dy}_{PW_{RD}}+\underbrace{\frac{1}{u_b}\int _{0}^{h}(\widetilde{u}-u_b)\overline{V_a}dy}_{VW_{RD}},
 \label{eq:RDqw-internal-finally}
\end{eqnarray}
and
 \begin{eqnarray}
q_w&=&\underbrace{\frac{1}{u_b}\int _{0}^{h} \overline{q_y}\frac{\partial \widetilde{u}}{\partial y}dy}_{MH_{RD}}+\underbrace{\frac{1}{u_b}\int _{0}^{h}C_v\overline{\rho T^{ \prime \prime} v^{ \prime \prime}}\frac{\partial \widetilde{u}}{\partial y}dy}_{TH_{RD}}\nonumber \\
&+&\underbrace{\frac{1}{u_b}\int _{0}^{h}\overline{p v}\frac{\partial \widetilde{u}}{\partial y}dy}_{PWT_{RD}}+\underbrace{\frac{1}{u_b}\int _{0}^{h}\overline{-{u_i}\tau_{i2}}\frac{\partial \widetilde{u}}{\partial y}dy}_{VWT_{RD}}\nonumber\\
&+&\underbrace{\frac{1}{u_b}\int _{0}^{h}\overline{\rho v^{ \prime \prime} {u}_{k}^{ \prime \prime} {u}_{k}^{ \prime \prime}/2}\frac{\partial \widetilde{u}}{\partial y}dy}_{TTH_{RD}}+\underbrace{\frac{1}{u_b}\int _{0}^{h} \widetilde{u_k}\overline{\rho v^{\prime \prime}{u_k}^{\prime \prime}}\frac{\partial \widetilde{u}}{\partial y}dy}_{RSH_{RD}}\nonumber\\
&+&\underbrace{\frac{1}{u_b}\int _{0}^{h} (u_b-\widetilde{u})(-\overline{fu})dy}_{EW_{RD}}.
 \label{eq:RDqw-energy-finally}
\end{eqnarray}

\begin{table*}
 \caption{Parameters on the three DNS cases: the flow conditions, the computational domains and the grid resolutions.}
  \label{tab:1}
  \begin{center}
\def~{\hphantom{0}}
  \begin{tabular}{ccccc c ccccc}
  \hline
case &$Re$ &$Re_\tau$ &$Ma$	 &$N_{x}$ &$N_{y}$ &$N_{z}$ & $\Delta x^+$ & $\Delta z^+$ & $\Delta y^+_{min}$ & $\Delta y^+_{max}$	\\ [3pt]
\hline
Ma05 &6000	 &355  &0.5  &400 &180 &320 &11.15 &4.64 &0.59 &8.23   \\
Ma15 &6000	 &408  &1.5  &400 &180 &320 &12.82 &5.34 &0.68 &10.68  \\
Ma30 &4880	 &456  &3.0	 &400 &210 &320 &14.32 &5.96 &0.65 &9.05   \\
   \hline
  \end{tabular}

  \end{center}
\end{table*}

\begin{table*}
\caption{Contributions to $B_q$ from different cases following equation~\eqref{eq:qw-e-mean}. $B_{q,integral}$ is the summation of the four integrals in equation~\eqref{eq:qw-e-mean}~\cite{zhangpeng2020}. }
  \label{tab:2}
  \begin{center}
\def~{\hphantom{0}}
 \scalebox{0.9}{
  \begin{tabular}{cccccccc}
  \hline
  case	  &$B_{q,direct}$	             &$B_{q,integral}$ 	               &$error$      &$TH$	                         &$MH$	                 &$PW$    	                  &$VW$    \\[3pt]
  \hline
  Ma05  &$-5.702\times10^{-3}$	 &$-5.758\times10^{-3}$  &0.98\% &$-3.057\times10^{-4}$ &$-1.793\times10^{-4}$	                                                            &$-1.060\times10^{-4}$	 &$-5.167\times10^{-3}$  \\
    &------	    &------   &------  &5.310\%	&3.114\%	 &1.841\%	   &89.735\%            \\
  Ma15	 &$-4.512\times10^{-2}$  &$-4.470\times10^{-2}$	   &0.93\%     &$-2.219\times10^{-3}$  &$-1.516\times10^{-3}$	                                                       &$-7.305\times10^{-4}$	 &$-4.023\times10^{-2}$ \\
   &------	   &------    &------  & 4.964\%	&3.393\%	  &1.634\%	    &90.009\%  \\
  Ma30  &$-1.392\times10^{-1}$  &$-1.405\times10^{-1}$    &0.93\%     & $-6.618\times10^{-3}$ &$-7.502\times10^{-3}$                                                            &$-2.051\times10^{-3}$	 & $-1.247\times10^{-1}$ \\
   &------	   &------    &------  &4.392\%	&5.341\%	 &1.460\%	 &88.807\%  \\
  \hline
  \end{tabular}}
  \end{center}
\end{table*}

\begin{table*}
\caption{Contributions to $B_q$ from different cases following equation~\eqref{eq:qw-E-mean}. $B_{q,integral}$ is the summation of the seven integrals in equation~\eqref{eq:qw-E-mean}.}
  \label{tab:3}
  \begin{center}
\def~{\hphantom{0}}
 \scalebox{0.7}{
  \begin{tabular}{ccccccccccc}
  \hline
  case	  &$B_{q,direct}$	 &$B_{q,integral}$ 	 &$error$ &$TH$	 &$MH$	 &$PWT$    	 &$VWT$     &$TTH$   &$RSH$ &$EW$               \\[3pt]
  \hline
  Ma05  &$-5.702\times10^{-3}$	 &$-5.786\times10^{-3}$      &1.47\% &$-3.057\times10^{-4}$ &$-1.793\times10^{-4}$	        &$-1.260\times10^{-4}$	 &$-1.965\times10^{-4}$                                                                                                            &$1.015\times10^{-4}$     &$-2.432\times10^{-3}$  &$-2.648\times10^{-3}$\\
    &------	    &------   &------  &5.284\%	&3.099\%	 &2.177\%	   &3.397\%    &$-1.755\%$      &42.037\%    &45.761\%    \\

  Ma15	 &$-4.512\times10^{-2}$  &$-4.509\times10^{-2}$	   &0.04\%     &$-2.219\times10^{-3}$  &$-1.516\times10^{-3}$	 &$-8.855\times10^{-4}$	 &$-1.829\times10^{-3}$   &$8.828\times10^{-4}$   &$-1.893\times10^{-2}$                                                                                                    &$-2.060\times10^{-2}$          \\
   &------	   &------    &------  & 4.920\%	&3.363\%	&1.964\%	  &4.056\%    &$-1.958\%$   &41.968\%
                                                                                       &45.687\%                \\

  Ma30  &$-1.392\times10^{-1}$  &$-1.403\times10^{-1}$    &0.77\%     & $-6.168\times10^{-3}$ &$-7.502\times10^{-3}$  &$-2.400\times10^{-3}$	 & $-1.009\times10^{-2}$   & $3.056\times10^{-3}$    & $-5.368\times10^{-2}$   & $-6.351\times10^{-2}$                \\
   &------	   &------    &------  &4.397\%	&5.348\%	 &1.710\%	 &7.195\%    &$-2.179\%$   &38.264\%    &45.265\%    \\
  \hline
  \end{tabular}}
  \end{center}
\end{table*}

\begin{table*}
\caption{Contributions to $B_q$ from different cases following equation~\eqref{eq:RDqw-internal-finally}. $B_{q,integral}$ is the summation of the four integrals from equation~\eqref{eq:RDqw-internal-finally}.}
  \label{tab:RD-internal}
  \begin{center}
\def~{\hphantom{0}}
 \scalebox{0.9}{
  \begin{tabular}{cccccccc}
  \hline
  case	  &$B_{q,direct}$	             &$B_{q,integral}$ 	               &$error$      &$TH_{RD}$	                         &$MH_{RD}$	                 &$PW_{RD}$    	                  &$VW_{RD}$    \\[3pt]
  \hline
  Ma05  &$-5.702\times10^{-3}$	 &$-5.677\times10^{-3}$  &0.44\%               &$-2.188\times10^{-3}$   &$-5.345\times10^{-4}$	                                               &$-1.893\times10^{-4}$	  &$-2.765\times10^{-3}$  \\
    &------	    &------   &------  &38.542\%	&9.415\%	 &3.335\%	   &48.708\%  \\
  Ma15	&$-4.512\times10^{-2}$  &$-4.515\times10^{-2}$  &0.066\%
  &$-1.791\times10^{-2}$  &$-3.924\times10^{-3}$	                                             &$-1.363\times10^{-3}$	 &$-2.195\times10^{-2}$ \\
   &------	   &------    &------  & 39.665\%	&8.691\%	 &3.020\%	   &48.625\%  \\
  Ma30  &$-1.392\times10^{-1}$  &$-1.403\times10^{-1}$  &0.79\%
  &$-6.016\times10^{-2}$  &$-9.869\times10^{-3}$                                                  &$-3.404\times10^{-3}$	 &$-6.684\times10^{-2}$ \\
   &------	   &------    &------  &42.890\%	&7.036\%	 &2.427\%	   &47.648\%  \\
  \hline
  \end{tabular}}
  \end{center}
\end{table*}

\begin{table*}
\caption{Contribution to $B_q$ from different cases following equation~\eqref{eq:RDqw-energy-finally}. $B_{q,integral}$is the summation of the seven integrals from equation~\eqref{eq:RDqw-energy-finally}.}
  \label{tab:RD-energy}
  \begin{center}
\def~{\hphantom{0}}
 \scalebox{0.7}{
  \begin{tabular}{ccccccccccc}
  \hline
  case	  &$B_{q,direct}$	 &$B_{q,integral}$ 	 &$error$ &$TH_{RD}$	 &$MH_{RD}$	 &$PWT_{RD}$    	 &$VWT_{RD}$     &$TTH_{RD}$   &$RSH_{RD}$ &$EW_{RD}$               \\[3pt]
  \hline
  Ma05  &$-5.702\times10^{-3}$	 &$-5.707\times10^{-3}$      &0.088\%
  &$-2.188\times10^{-3}$  &$-5.345\times10^{-4}$  &$-1.013\times10^{-3}$
  &$-2.209\times10^{-4}$  &$5.939\times10^{-5}$   &$-1.988\times10^{-3}$  &$1.774\times10^{-4}$\\
    &------	    &------   &------  &38.337\%	&9.365\%	 &17.742\%	   &3.871\%    &-1.041\%	 &34.834\%	 &-3.108\%\\
  Ma15	&$-4.512\times10^{-2}$  &$-4.527\times10^{-2}$	     &0.33\%
  &$-1.791\times10^{-2}$  &$-3.924\times10^{-3}$	&$-8.961\times10^{-3}$	 &$-1.565\times10^{-3}$  &$5.071\times10^{-4}$     &$-1.529\times10^{-2}$  &$1.869\times10^{-3}$ \\
   &------	   &------    &------  & 39.559\%	&8.667\%	 &19.795\%	   &3.456\%     &-1.120\%  &33.773\%    &-4.129\%\\
  Ma30   &$-1.392\times10^{-1}$  &$-1.406\times10^{-1}$      &1.01\%
  &$-6.106\times10^{-2}$ &$-9.869\times10^{-3}$   &$-3.595\times10^{-2}$	 &$-3.789\times10^{-3}$ &$1.530\times10^{-3}$    &$-4.245\times10^{-2}$
  &$1.006\times10^{-2}$ \\
   &------	   &------    &------  &42.781\%	&7.018\%	 &25.560\%	   &2.694\%     &-1.088\%  &30.188\%    &-7.154\%\\
  \hline
  \end{tabular}}
  \end{center}
\end{table*}

\noindent Equations \eqref{eq:RDqw-internal-finally} and \eqref{eq:RDqw-energy-finally} are two new exact mathematical formulas for $q_w$ for compressible channel flows with symmetric isothermal wall boundary conditions for temperature and no-slip wall boundary conditions for all three velocity components. If the boundary conditions are changed, there might be some extra terms, such as $[(\widetilde{u}-u_b)\overline{ev}](y=h)-[(\widetilde{u}-u_b)\overline{ev}](y=0)$. It is interesting to see that the formulas for $q_w$ by two different methods are quite similar in the expressions, i.e. equation~\eqref{eq:qw-e-mean} versus equation~\eqref{eq:RDqw-internal-finally} and equation~\eqref{eq:qw-E-mean} versus equation~\eqref{eq:RDqw-energy-finally}. If we substituted $(\widetilde{u}-u_b)$ and $1/u_b$ in equations~\eqref{eq:RDqw-internal-finally} and \eqref{eq:RDqw-energy-finally} by $(y-h)$ and $1/h$, equations~\eqref{eq:qw-e-mean} and~\eqref{eq:qw-E-mean} with $\delta=h$ would be recovered, since
 \begin{eqnarray}
\underbrace{ \int_0^h\overline{q_y}dy=\int_0^h\overline{q_y}\frac{\partial (y-h)}{\partial y}dy}_{ y-derivative-weighting~ in~ Eq.\eqref{eq:qw-e-mean}}, \nonumber\\
 \underbrace{ \int _{0}^{h} \overline{q_y}\frac{\partial (\widetilde{u}-u_b)}{\partial y}dy =\int _{0}^{h} \overline{q_y}\frac{\partial \widetilde{u}}{\partial y}dy}_{u-derivative-weighting~ in~ Eq.\eqref{eq:RDqw-internal-finally}},\nonumber
\end{eqnarray}
for example.
\noindent Therefore, we may conclude that the different formulas derived by the FIK and RD methods are essentially different in the weight parameters. The former ones are $(y-h)$ and $1/h$ (distance related), while the latter ones are $(\widetilde{u}-u_b)$ and $1/u_b$ (velocity related). Due to the similarity of the equations, the name of the terms in equations \eqref{eq:RDqw-internal-finally} and \eqref{eq:RDqw-energy-finally} are denoted with the same ones as those in equations \eqref{eq:qw-e-mean} and \eqref{eq:qw-E-mean} but with a subscript "RD" to show the difference.

\section{Numerical verifications}

In this section, the direct numerical simulations (DNS) data of compressible turbulent channel flows with symmetric isothermal boundary conditions under different Reynolds and Mach numbers will be used to validate the above mathematical formulas. The DNS data are the same as those used by Zhang and Xia~\cite{zhangpeng2020}, and the parameters of the three cases are listed in table~\ref{tab:1}. For more information about the DNS details, the readers are referred to the paper by Zhang and Xia~\cite{zhangpeng2020}. In the following data, the wall heat flux coefficient $B_q=q_w/(\rho_w C_p u_\tau T_w)$ will be used to quantify the wall heat flux, where $1/(\rho_w C_p u_\tau T_w)=1.691$, 13.243 and 38.475 respectively for the three cases.

Tables \ref{tab:2} and \ref{tab:3} display the absolute and relative contributions to the wall heat flux coefficient $B_{q,integral}$ from the three DNS cases following equations~\eqref{eq:qw-e-mean} and \eqref{eq:qw-E-mean} with $\delta=h$, respectively, while tables~\ref{tab:RD-internal} and~\ref{tab:RD-energy} show the absolute and relative contributions to $B_{q,integral}$ from the three DNS cases following equations~\eqref{eq:RDqw-internal-finally} and \eqref{eq:RDqw-energy-finally}, respectively. Besides the integral results, the direct estimation of the wall heat flux coefficient $B_{q,direct}$ are also listed, where $B_{q,direct}$ is calculated directly through the wall heat flux using the mean temperature gradient at the wall. The relative error is calculated as
\begin{equation}
error=\left|\frac{B_{q,direct} -B_{q,integral}}{B_{q,direct}}\right|\times100\%.
\label{eq:error}
\end{equation}
From the errors listed in tables~\ref{tab:2}-~\ref{tab:RD-energy}, it is easy to see that all the four mathematical formulas, i.e. equations~\eqref{eq:qw-e-mean}-\eqref{eq:RDqw-energy-finally}, are ``exact'', and the relative errors between the integral formula and the direct estimation are within $1.5\%$ for all cases.

According to the results listed in table~\ref{tab:2}, the viscous work term VW dominantly contributes to $B_q$, approximately $90\%$, following equation~\eqref{eq:qw-e-mean}, and around $90\%$ of VW contribution comes from the near wall region with $y/h \leq 0.2$ according to the cumulative contribution analysis with different wall-height locations~\cite{zhangpeng2020}, which is consistent with the physical intuition. According to equation~\eqref{eq:qw-E-mean} and the data listed in table~\ref{tab:3}, the main contributions to $q_w$ are from the RSH and EW terms, i.e. the Reynolds stress transport term (around $40\%$) and the external-force work term (around $45\%$). The PWT, VWT and TTH terms contribute much less to $B_q$, and the first two terms have positive contributions while the third term, TTH, contributes negatively. Further discussions between equations~\eqref{eq:qw-e-mean} and \eqref{eq:qw-E-mean} will be left in the next section.

The data listed in tables~\ref{tab:RD-internal} and~\ref{tab:RD-energy} demonstrate that the formulas derived following the RD method, i.e. equations~\eqref{eq:RDqw-internal-finally} and \eqref{eq:RDqw-energy-finally}, display different results. According to equation~\eqref{eq:RDqw-internal-finally}, the main contribution to $B_q$ are the turbulent heat transfer term $TH_{RD}$ (around $40\%$) and the viscous stress work term $VW_{RD}$ (around $48\%$). That is, the relative contribution from the viscous stress work was become less for around $40\%$ if the RD method was used, as compared to those derived by the FIK method. From the results estimated from equation~\eqref{eq:RDqw-energy-finally}, the pressure work term and the Reynolds stress transport term contribute a lot to $B_q$ besides the turbulent heat transfer term. One striking result is that the contribution to $B_q$ from the external-force work is negative. The above results then demonstrate that the four exact mathematical formulas deliver different information on the contributions to $B_q$.

\section{Discussions}

\subsection{The total energy equation versus the internal energy equation}

\begin{figure}
\centering \includegraphics[scale=0.25]{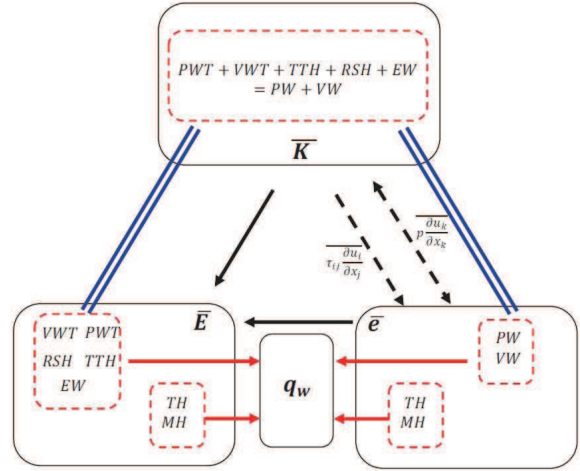}
\caption{Schematic diagram of energy conservation (black lines) among the mean total energy $\overline{E}$, the mean kinetic energy $\overline{K}$ and the mean internal energy $\overline{e}$ and the contribution terms (red lines) to $q_w$ from $\overline{E}$ and $\overline{e}$. $\overline{K}$ is not directly related to $q_w$, but acting as a bridge to connect the contribution terms to $q_w$ from $\overline{E}$ and $\overline{e}$ (blue lines).}
\label{fig:qw-energy}
\end{figure}

According to the mathematical formulas and results shown above, it is easy to see that the formulas obtained from the total energy equation and those obtained from the internal energy equations are quite different. These could be explained by the relation between the total energy equation and the internal energy equation. According to the definition, the total energy $E$ is the summation of the internal energy $e$ and the kinetic energy $K=\rho u_iu_i/2$, and thus the total energy equation is indeed connected with the internal energy equation through the kinetic energy equation. The pressure dilatation term $P_d$ and the viscous action term $V_a$ are in charge of the interactions between kinetic energy and internal energy. The former permits a two-way exchange while the latter can only lead to a one-way energy transfer from $K$ to $e$~\cite{Mittal2019}. A schematic diagram of energy conservation among the mean total energy $\overline{E}$, the mean kinetic energy $\overline{K}$ and the mean internal energy $\overline{e}$, and the contribution terms to $q_w$ from $\overline{E}$ and $\overline{e}$ is shown in figure~\ref{fig:qw-energy}. $\overline{K}$ is not directly related to $q_w$, but acting as a bridge to connect the contribution terms to $q_w$ from $\overline{E}$ and $\overline{e}$. The work done by the external force will directly contribute directly to $\overline{K}$ and $\overline{E}$, but not to $\overline{e}$, where its effect will be passed on through the interaction terms $P_d$ and $V_a$. The two contribution terms obtained by using the internal energy equation~\eqref{eq:e}, i.e. PW and VW, are mathematically equivalent to the five contribution terms derived from the total energy equation~\eqref{eq:E}, i.e. PWT, VWT, RSH, TTH and EW. It should be emphasized that the above mathematical statement is solid for the formulas derived following either the FIK method or the RD method. A mathematical derivation following the FIK method can be found in the Appendix.
From the data shown tables~\ref{tab:2} and~\ref{tab:3} and those shown in tables~\ref{tab:RD-internal} and~\ref{tab:RD-energy}, the above arguments can be easily verified. In summary, the wall heat flux formulas derived by using the internal energy equation (equations~\eqref{eq:qw-e-mean} and~\eqref{eq:RDqw-internal-finally}) and the total energy equation (equations~\eqref{eq:qw-E-mean} and~\eqref{eq:RDqw-energy-finally}) are indeed the same mathematically, but they represent different physical processes. The ones from the internal energy equation are more straightforward, while those from the total energy equation are more original and they can track all the energy input to the system and the work done by the external force.

\subsection{Arbitrary order of integrals for $q_w$ by the FIK method}

As discussed above, the FIK method is just the wall-normal integration of the balance equation. In this subsection, we will discuss about the arbitrary order of integrals for $q_w$ following the FIK method. Without loss of generality, the arbitrary order of integrals will be performed on the mean internal energy equation~\eqref{eq:e-mean} and the formulas for arbitrary order of integrals on the mean total energy equations~\eqref{eq:E-mean} can be derived in a similar way. The formula for symmetric compressible channel flows without blowing nor suction reads as

\begin{eqnarray}
q_w&=& \underbrace{ \frac  {n-1}{h^{n-1}} \int_{0}^{h} (h-y)^{n-2} \overline{q_y}  dy}_{MH}+ \underbrace{  \frac  {n-1} {h^{n-1}} \int_{0}^{h} (h-y)^{n-2} \overline{C_v \rho T ^{\prime \prime }  v ^{\prime \prime } } dy }_{TH} \nonumber\\
&+& \underbrace{ \frac  {1} {h^{n-1}} \int_0^{h} (h-y)^{n-1} (-\overline{P_d})dy}_{PW} \nonumber \\
&+& \underbrace{ \frac  {1} {h^{n-1}} \int_0^{h} (h-y)^{n-1} (-\overline{V_a})dy}_{VW},\quad\mathrm{for} \quad n\geq 2.
 \label{eq:balance2}
 \end{eqnarray}

\begin{table*}
\caption{The accuracy of equation~\eqref{eq:balance2} and the corresponding share of each term with five different $n$ for the case Ma15.}
\label{tab:6}
 \begin{center}
\def~{\hphantom{0}}

  \begin{tabular}{ccccccc}
  \hline
  integral time	  &$B_{q,integral}$        &$error$      &$TH$	        &$MH$	      &$PW$    	      &$VW$    \\[3pt]
  \hline
   n=2           &$-4.470\times10^{-2}$     & 0.93\%      & 4.964\%	&3.393\%	  &1.634\%	      &90.009\%  \\
   n=5           &$-4.493\times10^{-2}$    & 0.41\%      &9.395\%	    &11.739\%	  &3.250\%	      &75.616\%  \\
   n=10          &$-4.507\times10^{-2}$    & 0.12\%      & 10.414\%    &22.602\%	  &3.674\%	      &63.310\%  \\
   n=20          &$-4.515\times10^{-2}$   & 0.07\%      &9.005\%	    &38.041\%	  &3.184\%	      &49.769\%  \\
   n=50          &$-4.524\times10^{-2}$   & 0.26\%      &4.748\%	    &62.233\%	  &1.640\%	      &31.379\%  \\
  \hline
  \end{tabular}
  \end{center}
\end{table*}

\begin{table*}
\caption{The accuracy of equation~\eqref{eq:RDqw-internal-finally-random} and the corresponding share of each term with five different $\alpha$ for the case Ma15.}
\label{tab:RD-alpha}
 \begin{center}
\def~{\hphantom{0}}

  \begin{tabular}{ccccccc}
  \hline
  $\alpha$	  &$B_{q,integral}$        &$error$      &$TH_{RD}$	        &$MH_{RD}$	      &$PW_{RD}$    	      &$VW_{RD}$    \\[3pt]
  \hline
   $\alpha=1$           &$-4.515\times10^{-2}$    & 0.066\%     &39.665\%	    &8.691\%	  &3.020\%	      &48.625\%  \\
   $\alpha=5$           &$-4.464\times10^{-2}$    & 1.06\%      &8.022\%	    &1.758\%	  &0.648\%	      &89.573\%  \\
   $\alpha=10$          &$-4.458\times10^{-2}$    & 1.20\%      &4.017\%       &0.880\%	  &0.347\%	      &94.756\%  \\
   $\alpha=20$          &$-4.455\times10^{-2}$    & 1.26\%      &2.010\%	    &0.440\%	  &0.197\%	      &97.353\%  \\
   $\alpha=50$          &$-4.453\times10^{-2}$    & 1.31\%      &0.804\%	    &0.176\%	  &0.107\%	      &98.913\%  \\
  \hline
  \end{tabular}
  \end{center}
\end{table*}

\begin{figure}
\centering
\includegraphics[scale=0.75]{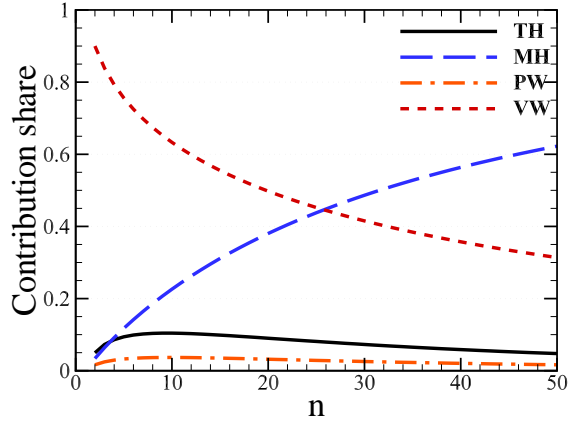}
\caption{Shares of the four contribution terms to $q_w$ in equation~\eqref{eq:balance2} with different integral order $n$ for the case M15.}
\label{fig:ntime}
\end{figure}

\noindent when $n=2$, the above equation~\eqref{eq:balance2} is exactly the equation~\eqref{eq:qw-e-mean} with $\delta=h$, which represents the mathematical averaged values in $[0, h]$ region of the balance terms in equation~\eqref{eq:qw-em1}. Table~\ref{tab:6} gives the accuracy of equation~\eqref{eq:balance2} and the corresponding contribution terms at five different integral order $n$ for the case M15. It is seen that the relative errors at different $n$ are all very small, less than $1\%$, documenting the high accuracy of the equation~\eqref{eq:balance2}. Figure~\ref{fig:ntime} shows the relative contribution share to $q_w$ of each term in equation~\eqref{eq:balance2} with different integral order $n$ for the case M15. From the results, it is found that with increasing $n$, the main contribution term to $q_w$ was changed from the VW term to the MH term, where the VW term contributes around $90\%$ of $q_w$ at $n=2$ while the MH term contributes around $62\%$ at $n=50$. The two other terms contribute bits of $q_w$, where the TH term is around $4\%-10\%$ and the PW term is around $1\%-4\%$ for $2\leq n \leq 50$. According to equation~\eqref{eq:balance2}, with increasing $n$, the contribution from the near wall region will become important since the weight factors $(h-y)^{n-2}$ and $(h-y)^{n-1}$ decay very fast with $y$ for $n>2$. The larger $n$ is, the faster the weight factors decay. For the results shown in Figure 4 and 5 in Zhang and Xia~\cite{zhangpeng2020}, $\overline{q_y}$ and $(1-y)\overline{V_a}$ have the largest value at the wall, while the peak value of $C_v\overline{\rho T''v''}$ and $(1-y)\overline{P_d}$ locates away from the wall. This could explain why the main contributions are from VW and MH as $n$ increases. On the other hand, there is a factor $(n-1)$ in the MH term, which could explain the increasing contribution share of the MH term over the VW term.

However, it should be emphasized that the equation~\eqref{eq:balance2} is just a mathematically strict formula, and we cannot find a reasonable physical interpretation for it when $n\geq3$. This is also the case for the skin-friction formula derived by Fukagata et al.~\cite{FIK2002}. In fact, Renard and Deck~\cite{DECK2016} also argued that the original FIK decomposition is of mathematical nature since variants of the FIK identity could be obtained by integrating the momentum equation four times instead of three.

\subsection{Arbitrary velocity-weight coefficient for $q_w$ by the RD method}

In the above derivations for equations~\eqref{eq:RDqw-internal-finally} and~\eqref{eq:RDqw-energy-finally}, a weight factor $(\widetilde{u}-u_b)$ was multiplied to the mean energy equation. In fact, the weight factor could be replaced by a more arbitrary form, such as $(\widetilde{u}-\alpha u_b)$, where $\alpha\neq 0$ is a parameter. For fully developed compressible channel flows with symmetric isothermal wall boundary conditions for temperature and no-slip wall boundary conditions for all three velocity components, we may arrive at the following formula:
 \begin{eqnarray}
 q_w=\underbrace{\frac{1}{ \alpha u_b}\int _{0}^{h} \overline{q_y}\frac{\partial \widetilde{u}}{\partial y}dy}_{MH_{RD}}+\underbrace{\frac{1}{ \alpha u_b}\int _{0}^{h}C_v\overline{\rho T^{ \prime \prime} v^{ \prime \prime}}\frac{\partial \widetilde{u}}{\partial y}dy}_{TH_{RD}}\nonumber\\
 +
\underbrace{\frac{1}{ \alpha u_b}\int _{0}^{h}(\widetilde{u}-\alpha u_b)\overline{P_d}dy}_{PW_{RD}}
+\underbrace{\frac{1}{ \alpha u_b}\int _{0}^{h}(\widetilde{u}-\alpha u_b)\overline{V_a}dy}_{VW_{RD}}.
 \label{eq:RDqw-internal-finally-random}
\end{eqnarray}

Again, if the boundary conditions are changed, there might be some extra terms, such as $[(\widetilde{u}-\alpha u_b)\overline{ev}](y=h)-[(\widetilde{u}-\alpha u_b)\overline{ev}](y=0)$. Mathematically, the above equation~\eqref{eq:RDqw-internal-finally-random} is mathematically correct for $\alpha\neq 0$. However, if $\alpha$ is very close to zero, the error of the above equation could be very large since there is factor $1/\alpha$. Table~\ref{tab:RD-alpha} shows the accuracy of equation~\eqref{eq:RDqw-internal-finally-random} and the corresponding share of each contribution term with five different $\alpha$ for the case Ma15. It is seen that equation~\eqref{eq:RDqw-internal-finally-random} is very accurate for five different $\alpha$, with a relative error less than $1.5\%$. Furthermore, it is seen that with increasing $\alpha$, the contribution share from the $TH_{RD}$ term decays rapidly while that from the $VW_{RD}$ term increases rapidly, becoming the dominant contribution term, as can also be seen from figure~\ref{fig:alpha}. The contribution from the molecular heat transfer ($MH_{RD}$), and the contribution from the pressure work ($PW_{RD}$) will also decay to zero as $\alpha$ increases. As $\alpha$ goes to infinity, equation~\eqref{eq:RDqw-internal-finally-random} can be simplified to:
\begin{figure}
\centering
\includegraphics[scale=0.75]{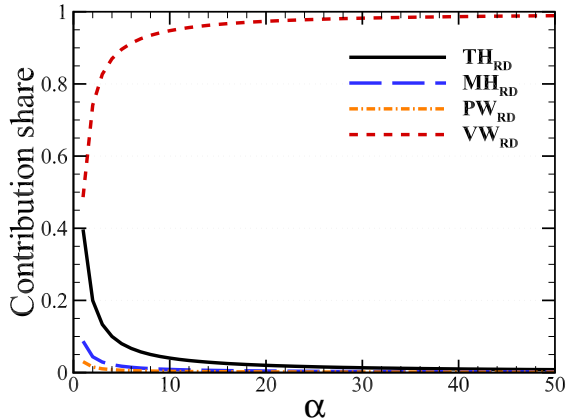}
\caption{Shares of the four contribution terms to $q_w$ in equation~\eqref{eq:RDqw-internal-finally-random} with different $\alpha$ for the case M15. }
\label{fig:alpha}
\end{figure}

\begin{eqnarray}
 q_w\approx-\int _{0}^{h}\overline{P_d}dy-\int _{0}^{h}\overline{V_a}dy,
 \label{eq:RDqw-infinity}
\end{eqnarray}
which is exactly the equation~\eqref{eq:qw-em1} with $y=h$. From this formula, the contribution due to the viscous stress work is also the dominant one, and its relative contribution share is even higher than the results obtained by equation~\eqref{eq:qw-e-mean}.

In the RD decomposition for the skin-friction, Renard and Deck~\cite{DECK2016} argued that the RD decomposition for the skin-friction has a straightforward physical interpretation since it relies on an energy budget, although the same formula can be derived through a short route as deduced in the above derivations. However, for the wall-heat flux discussed in the present work, the physical interpretation on equations~\eqref{eq:RDqw-internal-finally}, \eqref{eq:RDqw-energy-finally} and~\eqref{eq:RDqw-internal-finally-random} is not clear. All we could say is that these formulas are mathematically correct ones.

\subsection{Final remarks}

As discussed in section~\ref{sec:qw_FIK}, the energy balance equations~\eqref{eq:qw-em1} and~\eqref{eq:qw-Em1} can also be used to estimate $q_w$ at any wall-normal location $y$. For compressible channel flows with symmetric isothermal wall boundary conditions for temperature and no-slip wall boundary conditions for all three velocity components, the choice with $y=h$ makes the estimation very special since the mean values of the terms in equations~\eqref{eq:qw-em1} and~\eqref{eq:qw-Em1} are zero at the wall and at $y=h$ (due to the symmetry). Equation~\eqref{eq:qw-em1} becomes equation~\eqref{eq:RDqw-infinity}, indicating that the heat transfer out of the channel through the wall equals to the pressure-dilatation work and the viscous stress work across the channel. The former term contributes very little to $q_w$ while the latter term contributes more than $98\%$ of $q_w$. We may also found that the main contribution of the viscous stress work comes from the near wall region by using the cumulative contribution analysis, which is consistent with our former conclusion~\cite{zhangpeng2020} and the physical intuition. It was argued by Zhang and Xia~\cite{zhangpeng2020} that equation~\eqref{eq:qw-em1} might mislead the relative contributions to $q_w$ if $q_w$ was estimated at different $y$, as shown by Ghosh et al.~\cite{Ghosh2010}. However, equation~\eqref{eq:qw-em1} with $y=h$ indeed show the right physical picture. Equation~\eqref{eq:qw-Em1} will reduce to
\begin{equation}\label{eq:qw-ES}
q_w=-\int_0^h \overline{fu}(y_1)dy_1=-f\int_0^h \overline{u}(y_1)dy_1.
\end{equation}
From the energy point of view, equation~\eqref{eq:qw-ES} indicates that the heat transfer out of the wall in the channel equals to the work done by the external force across the channel. This formula is indeed closely related to equation~\eqref{eq:Huang-b} since the external force can be related to wall shear stress $\tau_w$ through the momentum balance analysis. Combining the information delivered by equation~\eqref{eq:qw-ES} and equation~\eqref{eq:RDqw-infinity}, it is easy to arrive at the following physical picture: The external force will do work to fluid in the channel, and this work will transfer to heat through the pressure dilation and the viscous stress action (the kinetic energy equation acts as the bridge), and all the heat generated will transfer out of the channel through the wall.

From the above discussions, we may say that equations~\eqref{eq:qw-ES} and~\eqref{eq:RDqw-infinity} can deliver the right flow physics on $q_w$, while equations~\eqref{eq:qw-e-mean} and~\eqref{eq:qw-E-mean} (with $\delta=h$) can show the contributions of the terms appeared in the energy balance equations in an average sense. Other equations are rather exactly mathematical ones, which are lack of clear physical interpretations. They can be used to estimate $q_w$, but might not be able to explore the related underlying physics. This is in sharp contrast to the skin friction decompositions proposed by Fukagata et al.~\cite{FIK2002} and Renard and Deck~\cite{DECK2016}. By performing the integration different times to the mean momentum equation, Fukagata et al.~\cite{FIK2002} argued that integrating the mean momentum equation once, twice and triple times can lead to the force balance equation, the mean velocity profile and the flow rate from the velocity profile, respectively, which can be viewed as kinds of physical interpretation. More surprisingly, the triple integration across the channel can give very simple expressions of the skin friction for turbulent channel flows, pipe flows and the boundary layer flows. On the other hand, the RD decomposition of the skin friction~\cite{DECK2016} relied on the energy budget and thus has a straightforward physical interpretation.

\section{Conclusion}\label{sec:Con}

We have derived the several expressions for the wall heat flux $q_w$ in compressible channel flows from the internal energy equation and the total energy equation by following the FIK method and the RD method, and confirmed the accuracy of the formulas by using the direct numerical simulation data. The formulas derived following the FIK method and those derived following the RD method are of the same form but with different weight parameters. The former one is distance related while the latter is velocity related. The formulas derived by using the internal energy equation and the total energy equation are indeed the same mathematically, and they can be connected by the Reynolds-averaged kinetic energy equation. For example, the summation of the PW and VW terms in equation~\eqref{eq:qw-e-mean} equals the summation of the PWT,VWT,TTH,RSH and EW terms in equation~\eqref{eq:qw-E-mean}. It can be easily proven that the above mathematical statement is solid for the formulas derived following either the FIK method or the RD method. Furthermore, it should be noted that all the formulas derived in the present paper are mathematical ones, which generally are lack of clear physical interpretation, except for equations~\eqref{eq:qw-ES},~\eqref{eq:RDqw-infinity},~\eqref{eq:qw-e-mean} and~\eqref{eq:qw-E-mean}. They can be used to estimate the wall-heat flux at the wall, but special attention should be paid when they were adopted to explore the related underlying physics.

\section*{Appendix} \label{app:1}

In this appendix, we will show that the summation of the PW and VW terms in equation~\eqref{eq:qw-e-mean} equals exactly the summation of the PWT, VWT, TTH, RSH and EW terms in equation~\eqref{eq:qw-E-mean}. This could be done by the analysis of the kinetic energy equation~\eqref{eq:kinetic energy} (This equation can be derived from the momentum equation or it can be obtained by subtracting the internal energy equation~\eqref{eq:E} from the total energy equation~\eqref{eq:e}):
\begin{equation}
\frac{\partial K}{\partial t}+\frac{\partial{K u_k}}{\partial x_k}+\frac{\partial\left( p u_k -\tau_{i k} u_i\right)}{\partial x_k}=p\frac{\partial u_k}{\partial x_k}-\tau_{i j}\frac{\partial u_i}{\partial x_j}+f u,
 \label{eq:kinetic energy}
\end{equation}

The Reynolds-averaged kinetic energy equation in compressible turbulence channel flows can then be obtained
\begin{eqnarray}
\frac{\partial\left(\widetilde{u_k} \overline{\rho v^{\prime \prime} u_k^{\prime \prime}  }+\overline{\rho v^{ \prime \prime} {u}_{k}^{ \prime \prime} {u}_{k}^{ \prime \prime}/2}+\overline{p v}-\overline{u_i \tau_{i2}} \right)}{\partial y}&=&\overline{p\frac{\partial u_k}{\partial x_k}}-\overline{\tau_{i j}\frac{\partial u_i}{\partial x_j}}\nonumber\\\
&+&\overline{f u},
 \label{eq:Reynolds-averaged kinetic energy}
\end{eqnarray}

By integrating equation~\eqref{eq:Reynolds-averaged kinetic energy} from 0 to $y$, and apply the no-slip boundary conditions for the velocity, we can obtain
\begin{eqnarray}
\overline{p v}(y)&-&\overline{u_i \tau_{i2}} (y)+\overline{\rho v^{ \prime \prime} {u}_{k}^{ \prime \prime} {u}_{k}^{ \prime \prime}/2} (y)+\widetilde{u_k}\overline{\rho v^{\prime \prime} u_k^{\prime \prime}} (y)-\int_{0}^{y}\overline{f u}(y_1)dy_1 \nonumber\\
&=&-\int_{0}^{y}\overline{P_d}(y_1)dy_1-\int_{0}^{y}\overline{V_a}(y_1)dy_1,
 \label{eq:Reynolds-averaged kinetic energy-once}
\end{eqnarray}

By integrating equation~\eqref{eq:Reynolds-averaged kinetic energy-once} from 0 to $\delta$ and then dividing by $\delta$, we may arrive at
\begin{eqnarray}
&&\underbrace{\frac{1}{\delta}\int_0^{\delta}\overline{pv} dy}_{PWT}+\underbrace{\frac{1}{\delta}\int_0^{\delta} -\overline{u_i \tau_{i2} }dy}_{VWT}+
\underbrace{\frac{1}{\delta}\int_0^{\delta} \overline{\rho v'' u_{k}''u_{k}''/2}dy}_{TTH} \nonumber\\
&+&\underbrace{\frac{1}{\delta}\int_0^{\delta} \widetilde{{u}_{k}} \overline{{\rho} {v}'' {u}_{{k}}''} dy}_{RSH}+
\underbrace{\frac{1}{\delta}\int_0^{\delta} (\delta-y) (-\overline{fu})dy}_{EW}\nonumber\\
&=&\underbrace{\frac{1}{\delta}\int_0^{\delta}(y-\delta)\overline{P_d} dy}_{PW} +\underbrace{\frac{1}{\delta}\int_0^{\delta} (y-\delta)\overline{V_a}dy}_{VW}.
 \label{eq:Reynolds-averaged kinetic energy-second}
\end{eqnarray}

We would like to thank the support by the National Natural Science Foundation of
China (NSFC grant nos. 11822208, 11772297, 91852205). P. Zhang and Y.B. Song contribute equally to the paper.

\bibliography{Ref}

\end{document}